\documentclass[aps,prc,twocolumn,superscriptaddress,showpacs]{revtex4}
\usepackage{epsfig}
\usepackage{amssymb}
\usepackage{graphicx}
\usepackage{dcolumn}
\newcommand{\beq}{\begin{equation}}
\newcommand{\eeq}{\end{equation}}
\newcommand{\beqar}{\begin{eqnarray}}
\newcommand{\eeqar}{\end{eqnarray}}
\newcommand{\ds}{\displaystyle}
\begin{document}

\title{Microscopic models and effective equation of state in 
    nuclear collisions at FAIR energies }

\author{L.V.~Bravina}
\altaffiliation[Also at ]{Skobeltzyn Institute for Nuclear Physics,
Moscow State University, RU-119899 Moscow, Russia
\vspace*{1ex}}
\affiliation{
Department of Physics, University of Oslo, PB 1048 Blindern,
N-0316 Oslo, Norway
\vspace*{1ex}}
\author{I.~Arsene}
\affiliation{
Department of Physics, University of Oslo, PB 1048 Blindern,
N-0316 Oslo, Norway
\vspace*{1ex}}
\author{J.~Bleibel}
\affiliation{
Institute for Theoretical Physics, University of T\"ubingen,
Auf der Morgenstelle 14, D-72076 T\"ubingen, Germany
\vspace*{1ex}}
\author{M.~Bleicher}
\affiliation{
Institute for Theoretical Physics, University of Frankfurt,
Max-von-Laue-Str. 1, D-60438 Frankfurt a.M., Germany
\vspace*{1ex}}
\author{G.~Burau}
\affiliation{
Institute for Theoretical Physics, University of Frankfurt,
Max-von-Laue-Str. 1, D-60438 Frankfurt a.M., Germany
\vspace*{1ex}}
\author{Amand Faessler}
\affiliation{
Institute for Theoretical Physics, University of T\"ubingen,
Auf der Morgenstelle 14, D-72076 T\"ubingen, Germany
\vspace*{1ex}}
\author{C.~Fuchs}
\affiliation{
Institute for Theoretical Physics, University of T\"ubingen,
Auf der Morgenstelle 14, D-72076 T\"ubingen, Germany
\vspace*{1ex}}
\author{M.S.~Nilsson}
\affiliation{
Department of Physics, University of Oslo, PB 1048 Blindern,
N-0316 Oslo, Norway
\vspace*{1ex}}
\author{H.~St\"ocker}
\altaffiliation[Also at ]{Frankfurt Institute for Advanced Studies 
(FIAS), University of Frankfurt,
Max-von-Laue-Str. 1, D-60438 Frankfurt a.M., Germany
\vspace*{1ex}}
\affiliation{Gesellschaft f\"ur Schwerionenforschung mbH,
Planckstra{\ss}e 1, D-64291 Darmstadt, Germany
\vspace*{1ex}}
\author{K.~Tywoniuk}
\affiliation{
Department of Physics, University of Oslo, PB 1048 Blindern,
N-0316 Oslo, Norway
\vspace*{1ex}}
\author{E.E.~Zabrodin}
\altaffiliation[Also at ]{Skobeltzyn Institute for Nuclear Physics,
Moscow State University, RU-119899 Moscow, Russia
\vspace*{1ex}}
\affiliation{
Department of Physics, University of Oslo, PB 1048 Blindern,
N-0316 Oslo, Norway
\vspace*{1ex}}

\date{\today}

\begin{abstract}
Two microscopic models, UrQMD and QGSM, were employed to study the 
formation of locally equilibrated hot and dense nuclear matter in
heavy-ion collisions at energies from 11.6{\it A} GeV to 160{\it A} 
GeV. Analysis was performed for the fixed central cubic cell of 
volume $V = 125$\,fm$^3$ and for the expanding cell which followed 
the growth of the central area with uniformly distributed energy.
To decide whether or not the equilibrium was reached, results of
the microscopic calculations were compared to that of the statistical
thermal model. Both dynamical models indicate that the state of
kinetic, thermal and chemical equilibrium is nearly approached at
any bombarding energy after a certain relaxation period. The higher
the energy, the shorter the relaxation time. Equation of state has
a simple linear dependence $P = a(\sqrt{s}) \varepsilon$, where 
$a \equiv c_s^2$ is the sound velocity squared. It varies from 
$0.12\pm 0.01$ at $E_{lab} = 11.6${\it A} GeV to 
$0.145 \pm 0.005$ at $E_{lab} = 160${\it A} GeV. Change of the slope 
in $a(\sqrt{s})$ behavior occurs at $E_{lab} = 40${\it A} GeV and can
be assigned to the transition from baryon-rich to meson-dominated
matter. The phase diagrams in the $T - \mu_{\rm B}$ plane show the
presence of kinks along the lines of constant entropy per baryon.
These kinks are linked to the inelastic (i.e. chemical) freeze-out 
in the system.
\end{abstract}

%\begin{keyword}
%Keywords: ultrarelativistic heavy-ion collisions; Monte Carlo
%transport model; equilibration of matter; equation of state;
%statistical model of hadron gas 
%\PACS 25.75.-q, 24.10.Lx, 24.10.Pa, 64.30.-t

%25.75.-q Relativistic heavy-ion collisions
%24.10.Lx Monte Carlo simulations (including hadron and parton
%         cascades and string breaking models)
%24.10.Pa Thermal and statistical models
%64.30.-t Equations of state of specific substances

%\end{keyword}
%\end{frontmatter}

\maketitle

\section{Introduction}
\label{sec1}

\begin{figure}[htb]
\includegraphics[scale=0.325]{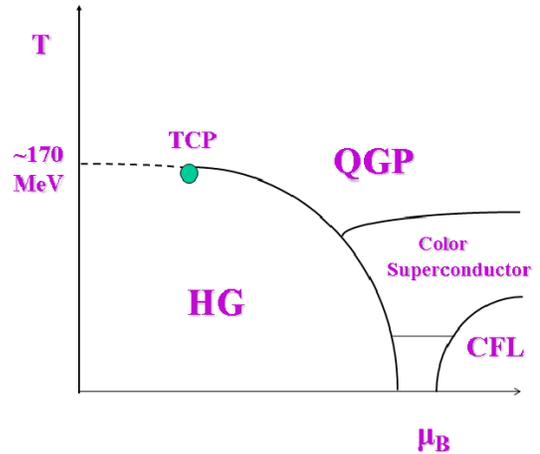}
\caption{
(Color online)
Present status of nuclear phase diagram in the $T-\mu_{\rm B}$ plane.
\label{fig1}}
\end{figure}

Experiments on heavy-ion collisions carried out for the last two
decades at GSI's Schwerionen Synchrotron (SIS), LBL's Bevalac, CERN's 
Super Proton Synchrotron (SPS), BNL's Alternating Gradient Synchrotron
(AGS) and Relativistic Heavy Ion Collider (RHIC) have significantly 
helped us in understanding of properties of hot and dense nuclear 
matter. The collisions at top RHIC energy $\sqrt{s} = 200${\it A} GeV 
or at energy of coming soon CERN's Large Hadron Collider (LHC)
$\sqrt{s} = 5.5${\it A} TeV probe the domain of high temperatures and 
low net baryon densities, while the systems with lower temperatures 
but with much higher baryon densities should be produced in heavy-ion 
collisions at relatively moderate energies around $E_{lab} = 
30${\it A} GeV accessible for future GSI's Facility for Antiproton 
and Ion Research (FAIR) accelerator \cite{fair}. Most likely, the 
matter under such extreme conditions is composed of partons, i.e. 
quarks and gluons, in the phase of quark-gluon plasma (QGP), colored 
tubes of chromo-electric field (or strings), hadrons and their 
resonances. The question about the equation of state (EOS) of such 
substance remains still open.  Present status of the nuclear phase 
diagram in terms of temperature $T$ and baryon chemical potential 
$\mu_B$ is sketched in Fig.~\ref{fig1}. 
The highly anticipated transition between the hot 
hadron gas (HG) and the QGP is of first order for relatively dense 
baryonic substances only. With rising temperature and dropping baryon 
density and chemical potential the transition becomes of second order 
at the so-called tricritical point (TCP). After that it is transformed
to a smooth crossover \cite{tcp}. Although the theory cannot localize 
the position of the TCP on $T$-$\mu_{\rm B}$ plane, lattice quantum 
chromodynamic (LQCD) calculations indicate that it might be somewhere 
between the points with $T \approx 160$\,MeV and $\mu_{\rm B} = 
360$\,MeV \cite{fodor} or $\mu_{\rm B} = 470\,$MeV \cite{karsch}. 
These values are close to the chemical freeze-out parameters obtained 
from the analysis of heavy-ion collisions at energies between 
$E_{lab} = 11.6${\it A} GeV and $E_{lab} = 40${\it A} GeV within the
statistical models \cite{ClRe99,ABS06,tsm_lu}. They are close also to 
the temperatures and baryon chemical potentials in the central zone 
of heavy-ion reactions generated by microscopic transport models 
\cite{plb98,LE1,jpg99,prc01,sqm06}. Another interesting feature of 
the collisions at bombarding energies around 40{\it A} GeV is the 
transition from baryon-dominated matter to meson-dominated one. 
According to microscopic models, in gold-gold collisions at top AGS 
energy nearly 70\% of total energy is deposited in baryonic sector. 
At top SPS energy mesons are carrying 70\% of the total energy, and 
at $E_{lab} \approx 40${\it A} GeV the energy parts of mesons and 
baryons are roughly the same. - The particle composition is changing. 
Is it possible to trace consequences of this change in microscopic 
model analysis? To answer this question two transport Monte Carlo 
models were employed: ultrarelativistic quantum molecular dynamics 
(UrQMD) model \cite{urqmd} and quark-gluon string model (QGSM) 
\cite{qgsm}. The models use different mechanisms of string excitation 
and fragmentation. UrQMD relies on the longitudinal excitation, while 
the color exchange scheme is employed in QGSM. Central gold-gold 
collisions with zero impact parameter $b = 0$\, fm were simulated at 
bombarding energies $E_{lab} = 11.6, 20, 30, 40, 80$ and 
160{\it A} GeV, respectively. Microscopic parameters related to 
quantities conserved in strong interactions, namely, the total energy, 
the net baryon charge, and the net strangeness extracted for a certain 
volume of the reaction were inserted into a system of nonlinear 
equations to obtain temperature, baryon chemical potential and 
strangeness chemical potential of an ideal hadron gas in equilibrium.
If the yields and transverse momentum spectra of particles obtained in
microscopic simulations are close to that provided by the statistical 
model, the matter in the cell is considered to be in the vicinity of 
equilibrium. Then its equation of state and other thermodynamic 
characteristics can be derived and studied.

Relaxation of hot matter to equilibrium in the central cell of central
heavy-ion collisions has been studied within the UrQMD model in
\cite{plb98,LE1,jpg99,prc01,sqm06} for energies ranging from 
11.6{\it A} GeV at AGS to $\sqrt{s} = 200${\it A} GeV at RHIC, and, 
partially, within the QGSM \cite{bleib_06,bleib_rec}. The size of the 
cell once chosen has been fixed throughout the system evolution. In 
the present paper we modify the analysis of the early stage in order 
to trace the expansion of an initially small area of homogeneity just 
after its formation. The central volume was further sub-separated 
into smaller cells embedded within each other ("matryoshka-doll" 
structure). The transition of analysis from the smaller cell to the 
larger one was allowed if, and only if, the energy densities in both 
cells were the same. Regardless of the microscopic model applied for 
the actual calculations, the formation of (quasi)equilibrated state in 
the central cell at all bombarding energies in question is observed. 
The matter in the cell expands isentropically with constant 
entropy-per-baryon ratio. The isentropic regime arises even before the 
chemical and thermal equilibration takes place. Due to coarse-graining 
of the central volume characteristic kinks in the temperature vs. 
baryochemical potential phase diagrams are found for both model 
simulations. This feature has not been seen in the previous studies 
because of the averaging of energy and baryon densities, in fact 
non-isotropically distributed within the relatively large volume. 

The paper is organized as follows. Similarities and differences of 
the microscopic string models chosen for the analysis are discussed in 
Sect.~\ref{sec2}. In Sect.~\ref{sec3} criteria of thermal and chemical 
equilibrium are formulated, and Sect.~\ref{sec4} describes the 
statistical model of an ideal hadron gas used for the comparison with 
both microscopic models. Section~\ref{sec5} presents the model study 
of the relaxation process in the cells with fixed and variable 
volumes. Special interest is paid to collisions at $E_{lab} = 
20${\it A} GeV and 40{\it A} GeV, and
obtained results are compared to that at neighbor energies. 
Conclusions are drawn in Sect.~\ref{sec6}.
  
\section{Features of UrQMD and QGSM}
\label{sec2}

\subsection{Similarities of the microscopic models}
\label{subsec2a}

Both UrQMD and QGSM are formulated as Monte-Carlo event generators
allowing to perform a careful analysis of the measurable quantities
by introducing  all necessary experimental cuts. The models are 
designed to describe hadronic, hadron-nucleus, and nuclear collisions
in a broad energy range. In the hadronic sector both models treat the 
production of new particles via formation and fragmentation of 
specific colored objects, strings. Strings are uniformly stretched, 
with constant string tension $\kappa \approx 1$\,GeV/fm,  between the 
quarks, diquarks and their antistates. The excited string is 
fragmenting into pieces via the Schwinger-like mechanism of 
$q\bar{q}$ and $q q - \bar{q} \bar{q}$ pair production, and the 
produced hadrons are uniformly distributed in the rapidity space. 

To describe hadron-nucleus $(hA)$ and nucleus-nucleus $(A+A)$ 
collisions the momenta and positions of nucleons in the nuclei are 
generated according to the Fermi momentum distribution and the 
Wood-Saxon density distribution, respectively. The black disk 
approximation is adopted as criterion of interaction. It means that 
two hadrons can interact both elastically and inelastically if the 
distance $d$ between them is smaller than $\sqrt{\sigma/\pi}$, where 
$\sigma$ is the total cross section. Tables of the experimentally 
available information, such as hadron cross sections, resonance 
widths and decay modes, are implemented in the models. If this 
information is lacking, the one-boson exchange model, detailed 
balance considerations and isospin symmetry conditions are employed. 
The propagation of particles is governed by Hamilton equation of 
motion, and both models use the concept of hadronic cascade for the 
description of $hA$ and $A+A$ interactions. Note that such a 
rescattering procedure is very important in the case of relativistic 
heavy-ion collisions and is necessary for the thermalization of the 
fireball. Due to the uncertainty principle newly produced 
particles can interact further only after a certain {\it formation
time\/}. However, hadrons containing the valence quarks can interact
immediately with the reduced cross section $\sigma = \sigma_{qN}$.
The Pauli principle is taken into account via the blocking of the 
final state, if the outgoing phase space is occupied. The Bose 
enhancement effects are not implemented yet.

\subsection{Differences between the microscopic models}
\label{subsec2b}

The differences between the models for hadronic interactions arise 
on three stages. First stage is the formation of strings. The UrQMD
belongs to group of models based on classical FRITIOF model 
\cite{fritiof}, while the QGSM uses the Gribov Reggeon field theory
(RFT) \cite{grt,PaSh07}. In the FRITIOF model the longitudinal 
excitation of
strings is employed, and the string masses arise from momentum
transfer. In the Gribov-Regge models the string masses appear due
to the color exchange mechanism, and strings are stretching between 
the constituents belonging to different hadrons. Longitudinal 
excitation of strings is also possible in the QGSM. This mechanism
describes the processes of single and double diffraction.
The second stage concerns the string fragmentation. The Lund JETSET
routine \cite{jetset}, used in the UrQMD, assumes that the string 
always breaks into a sub-string and a particle on a mass shell. In 
the QGSM the Field-Feynman algorithm \cite{FiFe} with independent
jets is applied. Therefore, the fragmentation functions which 
determine the energy, momentum, and the type of the hadrons produced
during the string decay, are different in the models.
The third stage deals with the number and type of the stings produced
in the collision. Due to the different mechanisms of string excitation 
and fragmentation, these numbers are also different for two 
microscopic models in question.
Last but not least, both models do not use the same tables of hadrons, 
chosen as discrete degrees of freedom. Whereas the UrQMD contains 55 
baryon and 32 meson states together with their antistates, the QGSM 
takes into account octet and decuplet baryons, and nonets of vector 
and pseudoscalar mesons, as well as their antiparticles. Further 
details can be found in \cite{urqmd} and \cite {qgsm}. 
Recently, the QGSM has been extended by the implementation of a 
parton recombination mechanism \cite{bleib_rec}. Since parton
recombination plays a minor role for nuclear collisions at 
intermediate energies, the whole analysis of the relaxation process 
is done for the standard QGSM.
We see that the basic underlying principles and designs of the models 
are quite far from each other. By using both the UrQMD and QGSM for 
studies of the relaxation process in a broad energy range one can 
expect that the model-dependent effects, caused by application of a 
particular event generator, will be significantly reduced.

\section{Statistical model of an ideal hadron gas}
\label{sec3}

For our analysis of the thermodynamic conditions in the cell we use
a conventional statistical model (SM) of an ideal hadron gas
formulated in pioneering works of Fermi \cite{Fer50} and Landau
\cite{La53}. The statistical approach was successfully applied to
the description of particle production in heavy-ion collisions 
from AGS to RHIC energies (see \cite{ABS06} and references therein).
In chemical and thermal equilibrium the distribution
functions of hadron species $i$ at temperature $T$ read (in units 
of $c = k_B = \hbar =1$)
\begin{equation}
\displaystyle
f(p,m_i) = \left[ \exp{ \left( \frac{\epsilon_i - \mu_i}{T} \right) }
\pm 1 \right] ^{-1}\ ,
\label{eq1}
\end{equation}
where $p$, $m_i$, $\epsilon_i = \sqrt{p^2 + m_i^2}$, and $\mu_i$ are
the full momentum, mass, energy, and the total chemical potential of
the hadron, respectively. The "$+$" sign is  for fermions 
and the "$-$" sign for bosons. Since in equilibrium the chemical
potentials associated to nonconserved charges vanish, the total 
chemical potential assigned to the $i$-th hadron is a linear 
combination of its baryon chemical potential $\mu_{\rm B}$ and 
strangeness chemical potential $\mu_{\rm S}$
\begin{equation}
\displaystyle
\mu_i = B_i \mu_{\rm B} + S_i \mu_{\rm S} \ ,
\label{eq2}
\end{equation}
with $B_i$ and $S_i$ being the baryon charge and the strangeness of
the particle, respectively. The isospin chemical potential (or,
alternatively, chemical potential associated with electric charge) 
is usually an order of magnitude weaker than $\mu_{\rm B}$ and
$\mu_{\rm S}$. Therefore, the dependence on this potential is 
disregarded in Eq.~(\ref{eq2}). Then, particle number density $n_i$ 
and energy density $\varepsilon_i$ are simply moments of the 
distribution function
\beqar
\ds
\label{eq3}
n_i &=&\frac{g_i}{(2\pi)^3}\int f(p,m_i) d^3p\ ,\\
\label{eq4}
\varepsilon_i &=& \frac{g_i}{(2\pi)^3}\int \sqrt{p^2+m_i^2}\, 
f(p,m_i) d^3p\ ,
\eeqar
with $g_i$ being the spin-isospin degeneracy factor of hadron $i$.
The partial hadron pressure given by the statistical model reads
\begin{equation}
\displaystyle
P_i = \frac{g_i}{(2\pi)^3}\int \frac{p^2}{3(p^2+m_i^2)^{1/2}} 
f(p,m_i) d^3p\ .
\label{eq5}
\end{equation} 

The integrals in Eqs.~(\ref{eq3})$-$(\ref{eq5}) can be calculated
numerically. Another way is to use a series expansion of 
Eq.~(\ref{eq1}) in the form \cite{La53}
\beq
\ds
f(p,m_i) = \sum \limits ^{\infty}_{n=1} (\mp 1)^{n+1} \exp{\left(- n
\frac{E_i - \mu_i}{T} \right) } \ ,
\label{eq6}
\eeq
which is inserted into Eqs.~(\ref{eq3})$-$(\ref{eq5}).  
After some straightforward calculations one gets
\beqar
\ds
\label{eq7}
n_i &=&\frac{g_i m_i^2 T}{2\pi^2}\sum_{n=1}^{\infty}
\frac{(\mp 1)^{n+1}}{n} \exp{ \left( \frac{n \mu_i}{T} \right)}
K_2\left(\frac{n m_i}{T} \right)\ , \\
\label{eq8}
\nonumber
\varepsilon_i &=& \frac{g_i m_i^2 T^2}{2\pi^2}\sum_{n=1}^{\infty}
\frac{(\mp 1)^{n+1}}{n^2} \exp{ \left( \frac{n \mu_i}{T} \right)} \\
&&\times \left[ 3 K_2\left(\frac{n m_i}{T} \right) + \frac{n m_i}{T}
K_1\left(\frac{n m_i}{T} \right) \right]\ , \\
\label{eq9}
P_i &=& \frac{g_i m_i^2 T^2}{2\pi^2}\sum_{n=1}^{\infty}
\frac{(\mp 1)^{n+1}}{n^2} \exp{ \left( \frac{n \mu_i}{T} \right)}
K_2\left(\frac{n m_i}{T} \right)\ ,
\eeqar
where $K_1$ and $K_2$ are modified Hankel functions of first and
second order, respectively. The first terms in 
Eqs.~(\ref{eq7})$-$(\ref{eq9}) correspond to the case of
Maxwell-Boltzmann statistics, which neglects the $\pm 1$ term in 
particle distribution function (\ref{eq1}).

The entropy density in the cell is represented by a sum over all 
particles of the product $f(p,m_i)\,[1 - \ln{f(p,m_i)}]$ integrated 
over all possible momentum states
\beq
\ds
s = -\sum_i \frac{g_i}{2\pi^2} \int_0^{\infty}
f(p,m_i)\, \left[ \ln{f(p,m_i)}-1 \right] \, p^2 d p.
\label{eq10}
\eeq

According to the presented formalism, the hadron composition and
energy spectra in equilibrium are determined by just three 
parameters, namely, the temperature, the baryon chemical potential,
and the strangeness chemical potential. In order to define values 
of $T,\ \mu_{\rm B}$, and $\mu_{\rm S}$ one
has to obtain the total energy density $\varepsilon$, baryon
density $\rho_{\rm B}$ and strangeness density $\rho_{\rm S}$ for
a given volume from microscopic model calculations, and insert them
as input parameters into the system of nonlinear equations 
\begin{eqnarray}
\displaystyle
\label{eq11}
\rho_B &=& \sum_i B_i\, n_i (T, \mu_{\rm B}, \mu_{\rm S})\ , \\ 
\label{eq12}
\rho_S &=& \sum_i S_i\, n_i (T, \mu_{\rm B}, \mu_{\rm S})\ ,  \\ 
\label{eq13}
\varepsilon &=& \sum_i \varepsilon_i (T, \mu_{\rm B}, \mu_{\rm S})\ ,
\end{eqnarray}
where $n_i (T, \mu_{\rm B}, \mu_{\rm S})$ and
$\varepsilon_i (T, \mu_{\rm B}, \mu_{\rm S})$ are given by 
Eqs.~(\ref{eq3})-(\ref{eq4}). Since the particle data tables 
implemented in the microscopic models contain different numbers of
hadrons, two versions of the SM with properly adjusted lists of 
hadron species are used, i.e. the number of hadronic degrees of
freedom in the macroscopic model should correspond to that in the 
microscopic model. To decide whether or not the equilibrium
is reached the criteria of the equilibrated state for open systems,
discussed in the next section, should be applied.

\section{Criteria of thermal and chemical equilibrium}
\label{sec4}

Criteria of local equilibrium for open systems were formulated in
\cite{LE1}, and we recall them briefly. Compared to a nonequilibrium
state, the equilibrium is characterized by the absence of collective 
effects, like flow of matter or flow of energy. The fireball produced 
in heavy-ion collisions is always expanding both radially and 
longitudinally. Therefore, the centrally placed symmetric cell is 
chosen to diminish effects caused by nonzero collective velocity of 
any asymmetric or asymmetrically located cell. The cell should be
neither too small to allow for the statistical treatment, nor too
large, - otherwise the homogeneous distribution of matter may not
be reached. Previous studies \cite{plb98,LE1,jpg99,prc01,bleib_06} 
found that the cubic cell of volume $V = 125$\, fm$^3$ centered 
around the center-of-mass of colliding gold-gold or lead-lead nuclei 
is well suited for such an analysis. Clearly, the relaxation to 
local equilibrium cannot occur earlier than at a certain time needed 
for the Lorentz contracted nuclei to pass through each other and 
leave the cell
\beq
\ds
t^{eq} \geq \frac{2R}{\gamma \beta} + \frac{\Delta z}{2 \beta}\ .
\label{eq14}
\eeq
Here $R$ is the nuclear radius, $\Delta z$ is the cell length in
longitudinal direction, $\beta$ is the velocity of nuclei in the
c.m. frame, and $\gamma = (1 - \beta^2)^{-1/2}$. Quite unexpectedly,
the reduction of the longitudinal size of the cell from 5\,fm to 
1\,fm does not automatically imply a faster equilibration in the 
smaller cell: the transition times are practically the same 
\cite{jpg99}. This means that the transition to equilibrium takes 
place simultaneously within a relatively large volume along the 
beam axis.  

\begin{figure}[htb]
\includegraphics[scale=0.525]{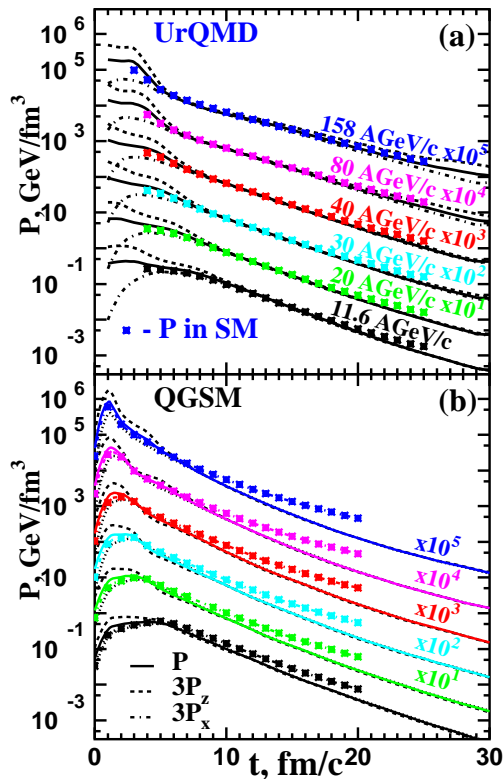}
\caption{
(Color online)
The longitudinal ($3 P_z$, dashed curves) and the
transverse ($3 P_x$, dash-dotted curves) diagonal components
of the microscopic pressure tensor in the central 125 fm$^3$ cell
in (a) UrQMD and (b) QGSM calculations of central Au+Au collisions
at energies from 11.6{\it A} GeV to 158{\it A} GeV. Asterisks
indicate the pressure given by the statistical model and solid
lines show the total microscopic pressure.
\label{p_vs_t}}
\end{figure}

Isotropy of the pressure gradients is a necessary condition for 
kinetic equilibration. Diagonal elements of the pressure tensor
$P_{\{x,y,z\}}$ are calculated from the virial theorem \cite{Bere92}
\beq
\ds
P_{\{x,y,z\}} = {1 \over {3V}}\, \sum_{i=h}
\frac{p^2_{i\{x,y,z\}}}{(m_i^2~+~p_i^2)^{1/2}}\ ,
\label{eq15}
\eeq
where $V$, $m_i$ and $p_i$ are the volume of the cell, the mass 
and the momentum of the $i$th hadron, respectively.
Figure~\ref{p_vs_t} depicts the convergence of the transverse 
pressure in the cell to the longitudinal one in the UrQMD and the 
QGSM calculations. Both models claim that the pressure becomes 
isotropic at $t \leq 10$\, fm/$c$ after beginning of the collision. 
The time of convergence decreases with rising bombarding energy. The
pressure calculated according to the statistical model is plotted 
onto the results of microscopic simulations also.
The agreement between microscopic and macroscopic calculations is
good for a period of about $ t = 8 - 10$\, fm/$c$. Then the matter
in the cell becomes quite dilute, and the collision rate is not 
sufficiently high to maintain equilibrium anymore. However, the
isotropy of pressure can be obtained, for instance, in a spherically 
expanding system of non-interacting particles. To exclude such a 
situation from the analysis one has to impose two additional 
criteria concerning thermal and chemical equilibrium. 

For a closed system in equilibrium the distribution functions of 
particles are given by Eq.~(\ref{eq1}) with an unique temperature, 
so the hadron composition and energy spectra are fixed. In open
systems neither the energy density nor the number of particles is 
conserved. Therefore, the snapshots of hadron abundances and energy 
spectra obtained at a certain time $t$ should be compared with
those corresponding to an ideal gas in equilibrium. The technical 
procedure is simple. At the very beginning, the pressure gradients
in transverse and longitudinal directions are considered. If the
pressure isotropy is restored, say, within 10\%-limit of accuracy,
the densities of conserved quantities, i.e. energy, baryon charge, 
and strangeness, determined microscopically, (i) should be used as an
input to Eqs.~(\ref{eq11})-(\ref{eq13}). The solution of this
system of equations (ii) provides us with values of the 
temperature, baryon chemical potential, and strangeness chemical 
potential which fully determine the composition and spectra of 
particles. By (iii) a comparison of microscopic and macroscopic 
yields of the most abundant hadronic species one can decide whether 
or not the chemical equilibrium occurs, whereas (iv) the
energy spectra of these hadrons should possess a common slope
corresponding to $1/T$ (thermal equilibrium). The similarity of the 
particle distributions means that our system is in the vicinity of
equilibrium. At each subsequent time step the procedure described
by (i)-(iv) is repeated. 

\section{Relaxation to equilibrium. Results and discussion}
\label{sec5}

\subsection{Yields and energy spectra}
\label{subsec5a}

\begin{figure}[htb]
\includegraphics[scale=0.425]{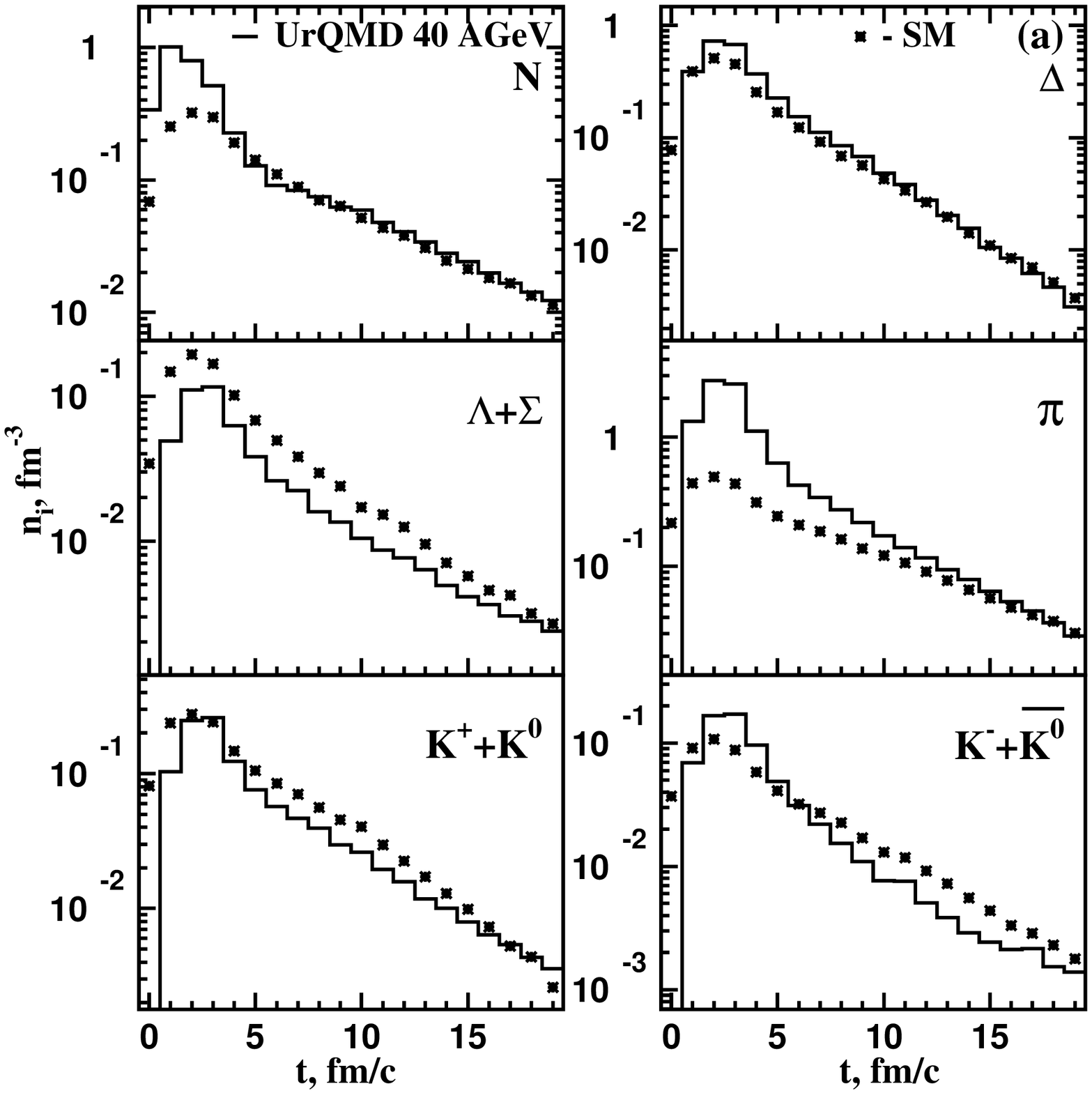}
\includegraphics[scale=0.425]{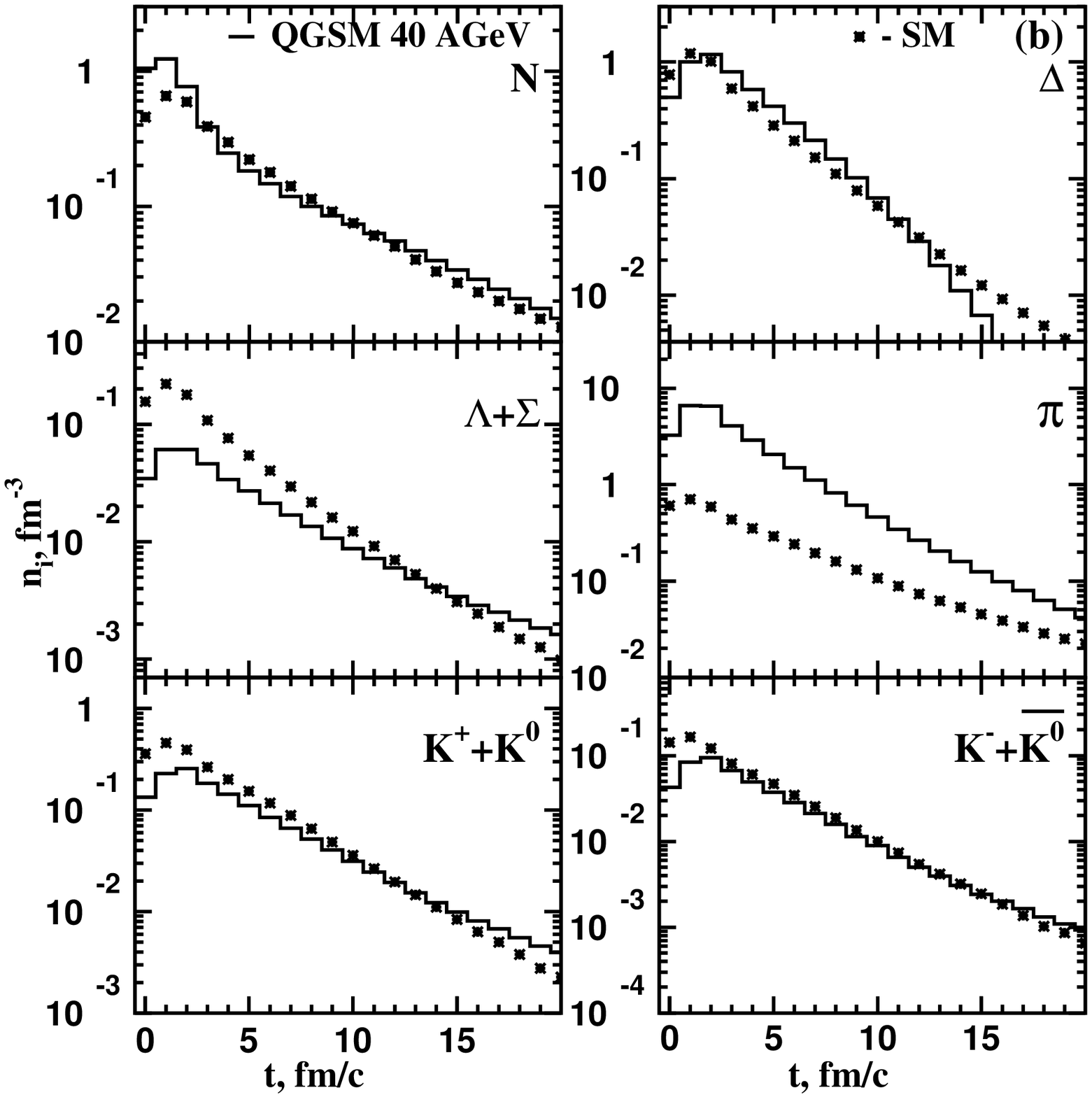}
\caption{
Evolution of yields of hadron species in the central cell
of volume $V = 125$\,fm$^3$ in (a) UrQMD and (b) QGSM calculations
(histograms) of central Au+Au collisions at 40{\it A} GeV. Asterisks
denote the results of the statistical model.
\label{yields}}
\end{figure}

The yields of main hadron species, i.e. $N, \Delta, \Lambda + 
\Sigma, \pi, K\ {\rm and}\ \overline{K}$ in the central cell are 
shown in Fig.~\ref{yields} for central Au+Au collisions at 
$E_{lab} = 40${\it A} GeV. For all particles, except pions, the 
agreement between the microscopic and macroscopic estimates at
$t \geq 9$\, fm/$c$ is good. Compared to the microscopic models,
the number of pions is underestimated in the SM. The pion excess
comes from the many-body decaying resonances, such as $N^{\ast},
\Delta^{\ast}, \Lambda^{\ast}, \omega$, etc, and strings. After
$t = 10-13$\, fm/$c$ the many-body processes play just a minor role,
and the pion multiplicity slowly converges to the equilibrium value.
It looks like all species of the hadronic cocktail, except pions, are
not far from the chemical equilibrium. It is well-known that the
pure statistical model of an ideal hadron gas, which does not include
effective chemical potential for pions or weak decays, systematically
underestimates the pion yields compared to experimental data.
Nevertheless, the excess of pions in the model with short table of
resonances, QGSM, is quite significant. This circumstance should 
affect the thermal spectra of all hadrons, provided the thermalization
is reached.   

\begin{figure}[htb]
\includegraphics[scale=0.525]{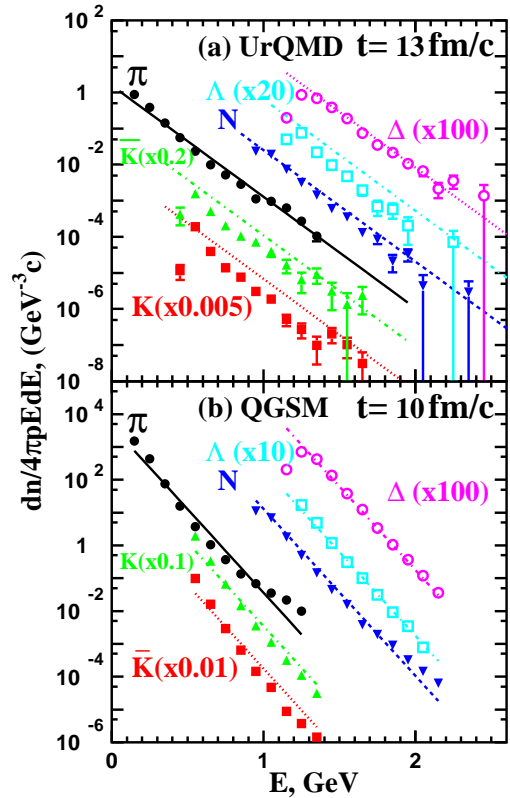}
\caption{
(Color online)
Energy spectra of $N$ ($\blacktriangledown $),
$\Lambda$ ($\square $), $\pi$ ($\bullet $), $\overline{K}$
($\blacktriangle $), $K$ ($\blacksquare$), and $\Delta$ ($\bigcirc$)
in the central 125 fm$^3$ cell
in (a) UrQMD and (b) QGSM calculations of central Au+Au collisions at
40{\it A} GeV at $t = 13$\,fm/$c$ and $t = 10$\,fm/$c$, respectively.
Lines show the results of the fit to Boltzmann distribution.
\label{en_spectra}}
\end{figure}

To verify how good the temperature is reproduced, the energy
spectra $d N / 4 \pi \rho E d E$ are displayed in
Fig.~\ref{en_spectra}. The Boltzmann fit to particle distributions 
is performed, and the SM calculations are plotted onto the 
microscopic results also. Both in UrQMD and in QGSM the energy
spectra agree well with the exponential form of the Boltzmann
distributions. Despite the good quality of the fit, the abundance of 
pions in particle spectrum leads to significant reduction of the 
effective temperature of the system within the QGSM calculations.
Analytical estimates of the temperature drop (see {\it Appendix\/}
\ref{app_A}) 
are close to the temperatures extracted from the fit. It would be 
possible to diminish the pion yield by taking into account larger part 
of the resonance states, but our intention is to check the principal 
occurrence of the (quasi)equilibrium states in different microscopic
models and to define the limits imposed on the effective equation of 
state. Note also, that significant part of the pion spectrum seems to 
be softer compared to other hadronic species. These pions are coming 
mainly from the decays of resonances and experience too few elastic 
collisions, that are necessary for their thermalization.
Since the hadronic matter in the central cell reaches the state of
thermal equilibrium, one can apply the mathematical apparatus 
formulated in Sect.~\ref{sec3} and, finally, obtain the anticipated
EOS.

\subsection{Evolution of the cell characteristics}
\label{subsec5b}

According to the information provided by Figs.~\ref{p_vs_t} -
\ref{en_spectra}, the appropriate time to start the study of
thermodynamic conditions in the cell is $t = 11$\,fm/$c$ for the
reactions at $E_{lab} = 20${\it A} GeV and $t = 9$\,fm/$c$ for 
$E_{lab} = 40${\it A} GeV. The input parameters obtained in the
microscopic model analysis are listed in Tables~\ref{tab1}-\ref{tab2} 
\begingroup
\squeezetable
\begin{table}
\caption{
\label{tab1}
The time evolution of the thermodynamic characteristics of hadronic
matter in the central cell of volume $V = 125$ fm$^3$ in central Au+Au
collisions at bombarding energy 20{\it A} GeV. The temperature, $T$,
baryochemical potential, $\mu_{\rm B}$, strange chemical potential,
$\mu_{\rm S}$, pressure, $P$, entropy density, $s$, and entropy
density per baryon, $s/\rho_{\rm B}$, are extracted from the
statistical model of ideal hadron gas, using the microscopically
evaluated energy density, $\varepsilon^{\rm cell}$, baryonic density,
$\rho_{\rm B}^{\rm cell}$, and strangeness density,
$\rho_{\rm S}^{\rm cell}$, as input. Of each pair of numbers, the
upper one corresponds to the UrQMD calculations, and the lower one to
the QGSM calculations.}
\vspace*{1ex}
\begin{ruledtabular}
\begin{tabular}{cccccccccc}
Time & $\varepsilon^{\rm cell}$ & $\rho_{\rm B}^{\rm cell}$ &
$\rho_{\rm S}^{\rm cell}$ & $T$ & $\mu_{\rm B}$ & $\mu_{\rm S}$ &
$P$ & $s$ & $s/\rho_{\rm B}^{\rm cell}$ \\
fm/$c$ & MeV/fm$^3$ & fm$^{-3}$ & fm$^{-3}$ & MeV & MeV & MeV &
 MeV/fm$^3$ & fm$^{-3}$ &  \\
\hline
11 & 464.2 & 0.210 & -0.0143 & 144.5 & 450.5 &  92.7 &  59.6 &
  2.97 & 14.16 \\
   & 522.6 & 0.257 & -0.0059 & 150.2 & 487.8 & 116.1 &  73.8 &
  3.13 & 12.19 \\
12 & 343.2 & 0.160 & -0.0115 & 137.9 & 459.2 &  86.4 &  44.0 &
  2.27 & 14.18 \\
   & 385.7 & 0.197 & -0.0051 & 141.9 & 498.1 & 109.4 &  53.1 &
  2.40 & 12.16 \\
13 & 255.2 & 0.124 & -0.0093 & 131.5 & 469.5 &  80.4 &  32.6 &
  1.75 & 14.15 \\
   & 286.9 & 0.153 & -0.0046 & 134.0 & 609.5 & 103.1 &  38.5 &
  1.85 & 12.09 \\
14 & 189.9 & 0.096 & -0.0072 & 124.9 & 481.7 &  75.8 &  24.1 &
  1.34 & 14.06 \\
   & 214.2 & 0.117 & -0.0035 & 127.2 & 515.9 &  97.1 &  28.2 &
  1.43 & 12.22 \\
15 & 143.9 & 0.075 & -0.0064 & 119.2 & 492.8 &  68.6 &  18.1 &
  1.05 & 13.97 \\
   & 162.3 & 0.091 & -0.0028 & 121.0 & 522.3 &  91.5 &  20.1 &
  1.12 & 12.35 \\
16 & 108.8 & 0.059 & -0.0052 & 113.7 & 502.5 &  62.7 &  13.6 &
  0.82 & 13.97 \\
   & 125.4 & 0.072 & -0.0025 & 115.4 & 529.2 &  85.4 &  15.9 &
  0.89 & 12.43 \\
17 &  83.6 & 0.046 & -0.0043 & 108.7 & 511.0 &  57.0 &  10.4 &
  0.65 & 14.02 \\
   &  98.3 & 0.058 & -0.0022 & 110.4 & 535.9 &  80.1 &  12.3 &
  0.72 & 12.52 \\
18 &  65.0 & 0.037 & -0.0035 & 103.5 & 523.7 &  52.4 &   8.0 &
  0.52 & 13.88 \\
   &  78.1 & 0.047 & -0.0019 & 105.9 & 541.3 &  75.4 &   9.6 &
  0.59 & 12.66 \\
19 &  50.9 & 0.030 & -0.0029 &  98.8 & 534.5 &  47.6 &   6.2 &
  0.41 & 13.82 \\
   &  62.9 & 0.039 & -0.0016 & 101.1 & 552.7 &  72.2 &   7.6 &
  0.49 & 12.52 \\
20 &  40.6 & 0.025 & -0.0027 &  94.6 & 544.2 &  38.9 &   4.8 &
  0.34 & 13.76 \\
   &  51.0 & 0.033 & -0.0014 &  97.0 & 560.1 &  67.4 &   6.0 &
  0.40 & 12.54 \\
\end{tabular}
\end{ruledtabular}
\end{table}
\endgroup
\begingroup
\squeezetable
\begin{table}
\caption{
\label{tab2}
The same as Table~\ref{tab1} but for 40{\it A} GeV.
}
\vspace*{1ex}
\begin{ruledtabular}
\begin{tabular}{cccccccccc}
Time & $\varepsilon^{\rm cell}$ & $\rho_{\rm B}^{\rm cell}$ &
$\rho_{\rm S}^{\rm cell}$ & $T$ & $\mu_{\rm B}$ & $\mu_{\rm S}$ &
$P$ & $s$ & $s/\rho_{\rm B}^{\rm cell}$ \\
fm/$c$ & MeV/fm$^3$ & fm$^{-3}$ & fm$^{-3}$ & MeV & MeV & MeV &
 MeV/fm$^3$ & fm$^{-3}$ &  \\
\hline
 9 & 662.3 & 0.226 & -0.0181 & 160.2 & 341.6 &  75.5 &  91.8 &
  4.23 & 18.69 \\
   & 732.3 & 0.290 & -0.0050 & 167.2 & 401.9 & 105.3 & 113.5 &
  4.36 & 15.01 \\
10 & 492.2 & 0.175 & -0.0145 & 153.2 & 354.2 &  71.6 &  68.3 &
  3.25 & 18.60 \\
   & 524.3 & 0.219 & -0.0041 & 157.1 & 417.9 & 100.4 &  79.3 &
  3.26 & 14.85 \\
11 & 369.4 & 0.135 & -0.0113 & 146.8 & 363.7 &  67.4 &  51.5 &
  2.53 & 18.73 \\
   & 384.5 & 0.170 & -0.0045 & 148.1 & 434.5 &  94.7 &  56.7 &
  2.48 & 14.61 \\
12 & 276.2 & 0.104 & -0.0094 & 140.5 & 374.7 &  62.4 &  38.7 &
  1.96 & 18.80 \\
   & 282.7 & 0.130 & -0.0033 & 140.0 & 447.2 &  90.0 &  40.9 &
  1.90 & 14.60 \\
13 & 205.7 & 0.081 & -0.0075 & 134.0 & 390.1 &  58.0 &  28.8 &
  1.51 & 18.66 \\
   & 211.5 & 0.101 & -0.0030 & 132.6 & 460.3 &  85.0 &  30.0 &
  1.47 & 14.53 \\
14 & 155.6 & 0.064 & -0.0060 & 128.0 & 404.0 &  54.9 &  21.8 &
  1.18 & 18.59 \\
   & 158.4 & 0.077 & -0.0023 & 126.3 & 465.9 &  79.4 &  22.2 &
  1.15 & 14.85 \\
15 & 118.9 & 0.050 & -0.0051 & 122.3 & 419.0 &  50.8 &  16.6 &
  0.93 & 18.43 \\
   & 120.4 & 0.060 & -0.0018 & 120.5 & 471.9 &  74.4 &  16.8 &
  0.90 & 15.16 \\
16 &  90.5 & 0.040 & -0.0041 & 117.2 & 426.6 &  46.3 &  12.8 &
  0.74 & 18.81 \\
   &  93.2 & 0.047 & -0.0013 & 115.2 & 479.7 &  71.2 &  12.9 &
  0.72 & 15.38 \\
17 &  69.9 & 0.032 & -0.0034 & 112.0 & 441.0 &  42.3 &   9.9 &
  0.59 & 18.69 \\
   &  73.8 & 0.038 & -0.0012 & 110.2 & 489.8 &  67.2 &  10.1 &
  0.59 & 15.39 \\
18 &  55.0 & 0.026 & -0.0028 & 107.0 & 457.3 &  39.2 &   7.6 &
  0.47 & 18.40 \\
   &  59.0 & 0.031 & -0.0006 & 105.7 & 499.7 &  70.0 &   7.9 &
  0.49 & 15.48 \\
19 &  43.3 & 0.021 & -0.0025 & 102.4 & 469.8 &  34.2 &   6.0 &
  0.39 & 18.34 \\
   &  47.8 & 0.026 & -0.0006 & 101.2 & 512.1 &  65.7 &   6.3 &
  0.40 & 15.31 \\
\end{tabular}
\end{ruledtabular}
\end{table}
\endgroup
together with the output thermodynamic characteristics given by the 
SM. Because of the different number of hadronic states employed by 
QGSM and UrQMD, the tables of available hadronic degrees of freedom 
in the statistical model are adjusted properly. The only objects not 
taken into account in the SM are strings. The detailed analysis 
done in \cite{box1,box2} shows that string processes play only a minor 
role at such late times in the central part of the reaction. Less
than 5\% of the total amount of hadronic collisions result to 
formation of strings. The strings produced at late time stages are 
quite light, and usually just one extra-particle, most commonly a 
pion, is produced after the string fragmentation. This circumstance, 
however, may account for the pion overproduction 
(see Fig.~\ref{yields}), since the inverse reactions such as 
3(or more)$\rightarrow$2 are not incorporated in the employed 
versions of both microscopic models.

\begin{figure}[htb]
\includegraphics[scale=0.525]{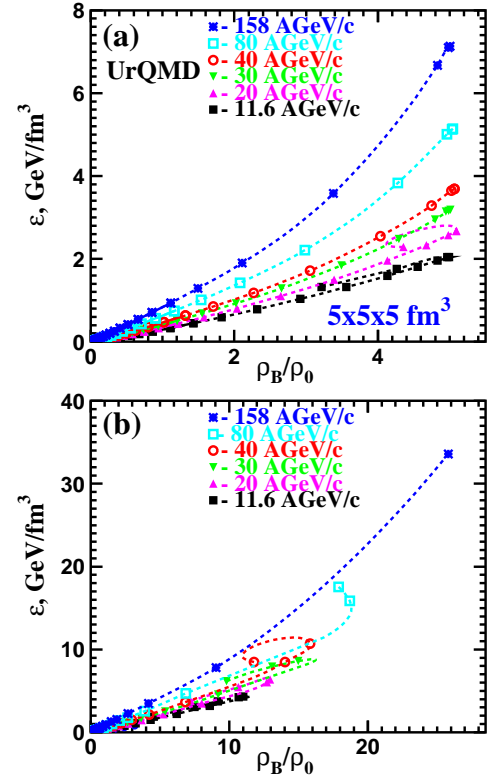}
\caption{
(Color online)
The total energy density $\varepsilon$ versus baryon density
$\rho_{\rm B}$ obtained in the central UrQMD cell of volume (a) $V =
125$\,fm$^3$ and (b) $V = 0.125$\,fm$^3$ during the time evolution of
central Au+Au collisions at energies from 11.6{\it A} GeV to
158{\it A} GeV. Dashed lines correspond to the non-equilibrium stage
of the reaction, solid lines represent the equilibrium phase.
\label{e_vs_rb_urqmd}}
\end{figure}

For both energies the baryon density in the cell at the beginning 
of the equilibrium phase is about 30\% larger than the normal baryon
density $\rho_0 = 0.16$\,fm$^{-3}$ in the UrQMD calculations. 
Whereas QGSM allows for the production of hot equilibrated matter with
a density of $\rho_{\rm B} = 1.8\,\rho_0$, much higher nuclear 
densities obtained in microscopic simulations have been reported 
\cite{cbm_dens}. One has to bear in mind two important things 
concerning such density estimates. Firstly, they are very sensitive 
to the volume of the test-system, especially at the initial stage of 
the collision. As seen in Fig.~\ref{e_vs_rb_urqmd}(a) and 
\begin{figure}[htb]
\includegraphics[scale=0.525]{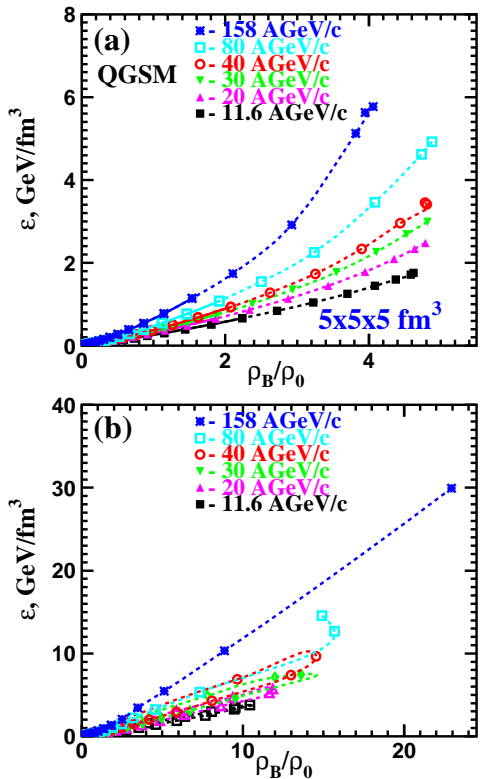}
\caption{
(Color online)
The same as Fig.~\protect\ref{e_vs_rb_urqmd}, but for QGSM
calculations.
\label{e_vs_rb_qgsm}}
\end{figure}
Fig.~\ref{e_vs_rb_qgsm}(a), the baryon density in both models cannot 
exceed $5\,\rho_0$ in the central cubic cell with volume 
$V = 5\times 5\times 5$\,fm$^3$ regardless of the bombarding energy, 
while for the smaller cell with volume $V_{small} = 0.5\times 0.5
\times 0.5$\, fm$^3$ the baryon density can be as high as 20\,$\rho_0$ 
in the calculations within the same microscopic models, see
Fig.~\ref{e_vs_rb_urqmd}(b) and Fig.~\ref{e_vs_rb_qgsm}(b). Secondly, 
such high values of the $\rho_{\rm B}$ should be treated with great 
care. The accelerated cold nuclear matter is automatically 
``compressed" in the calculations by the $\gamma$-factor. At the 
initial stage of a nuclear collision one deals with two opposite 
fluxes of Lorentz-contracted nucleons which just start to interact 
with their counterparts. Although the calculated baryon densities are 
huge, especially for the small cell, this is a purely kinematic effect, 
since the system is far from local equilibrium. The numbers become 
meaningful only when the equilibration takes place.

\begin{figure}[htb]
\includegraphics[scale=0.525]{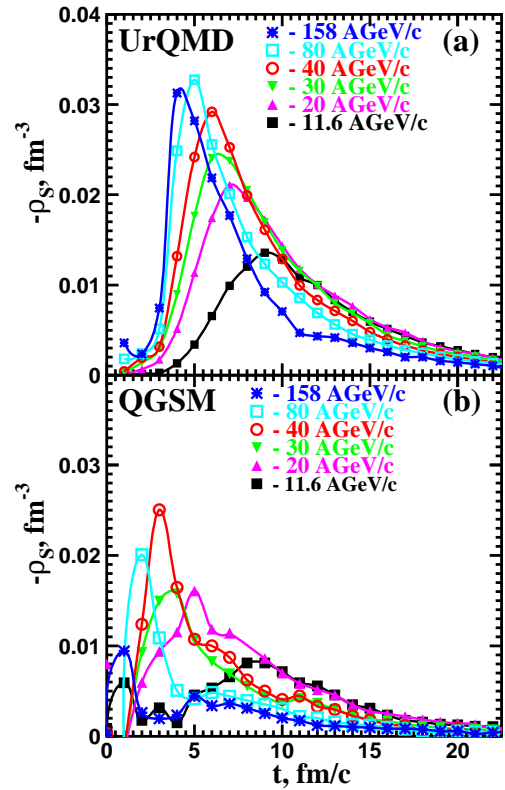}
\caption{
(Color online)
Time evolution of net strangeness density in the central
125 fm$^3$ cell in (a) UrQMD and (b) QGSM calculations of central
Au+Au collisions at energies from 11.6{\it A} GeV to 158{\it A} GeV.
Lines are drawn to guide the eye.
\label{rhos_vs_t}}
\end{figure}

Another interesting effect is the negative (though small) net 
strangeness density in the cell throughout the evolution of the
system depicted in Fig.~\ref{rhos_vs_t}. The result is pretty 
insensitive to the size of test-volume and can be explained as 
follows. Strange particles are always produced in pairs, so the net
$\rho_{\rm S}$ is zero. At energies about 40{\it A} GeV and below 
kaons emerge predominantly with lambdas and antikaons. Because of 
the $\bar{s}$-quark in its composition, kaons have significantly 
smaller interaction cross-section with baryons at $p \leq 
2$\,GeV/$c$ compared to antikaons, which carry the $s$-quark.
Therefore, $K$ leave the central cell with positive net baryon
charge easier than $\Lambda$ or $\overline{K}$ thus resulting to
negative net strangeness. At RHIC energies the ${\rm B} - 
\overline{\rm B}$ asymmetry in the cell is much less pronounced,
and the net $\rho_{\rm S}$ is very close to zero 
\cite{prc01,tsm_lu}.

\begin{figure}[htb]
\includegraphics[scale=0.525]{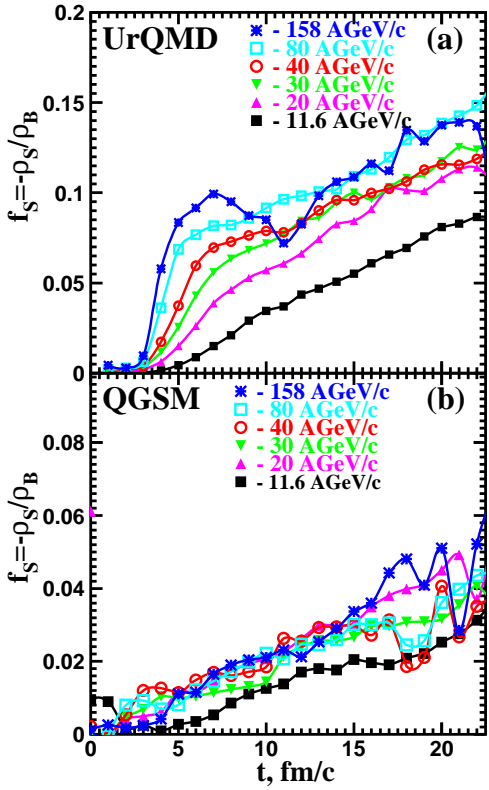}
\caption{
(Color online)
The same as Fig.~\protect\ref{rhos_vs_t} but for strangeness
per baryon, $f_s = - \rho_{\rm S} / \rho_{\rm B}$.
Lines are drawn to guide the eye.
\label{rsrb_vs_t}}
\end{figure}

Here we distinctly see the role of hadronic degrees of freedom. 
Despite the net baryon density is about 15\% larger in the QGSM
calculations than in the UrQMD ones, the absolute value of the net
strangeness density is almost 30\% higher in the UrQMD cell as 
compared to that in the QGSM. Extra-strangeness is deposited in the 
resonance sector, mainly in $\Lambda^\ast$ and $K^\ast$. Although 
the net $\rho_{\rm S}$ in the cell shown in Fig.~\ref{rhos_vs_t} 
quickly drops almost to zero after $t = 6$\,fm/$c$, its relaxation 
proceeds slower than that of the net baryon density. 
Figure~\ref{rsrb_vs_t} displays the instant rise of the ratio 
$f_s = - \rho_{\rm S} / \rho_{\rm B}$ with time $t$ attributed to
both microscopic models. The role of the small non-zero net 
strangeness is not negligible. The difference in particle spectra
and, especially, in particle ratios can be about 15\% \cite{LE1}
if one performs the SM calculations with essentially zero net
strangeness.        

\subsection{EOS in the cell}
\label{subsec5c}

\begin{figure}[htb]
\includegraphics[scale=0.525]{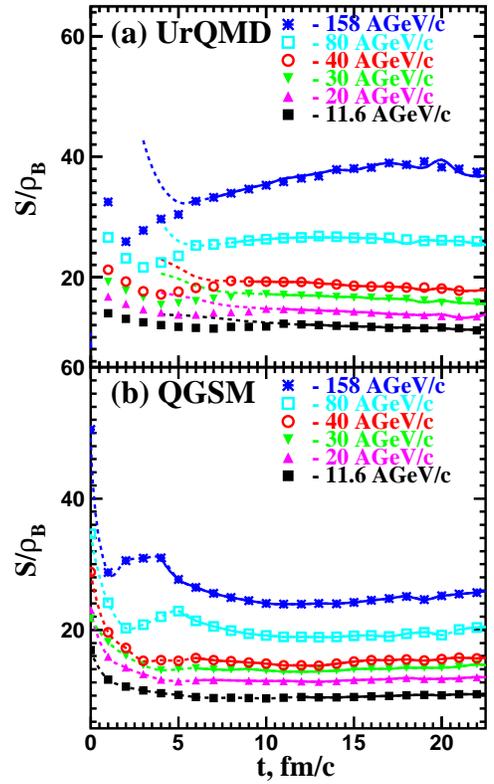}
\caption{
(Color online)
Time evolution of entropy per baryon $S / \rho_{\rm B}$ in
the central 125 fm$^3$ cell in (a) UrQMD and (b) QGSM calculations of
central Au+Au collisions at energies from 11.6{\it A} GeV to
158{\it A} GeV. Dashed lines correspond to the non-equilibrium stage
of the reaction, solid lines represent the equilibrium phase.
\label{s_vs_rhob}}
\end{figure}

Isentropic expansion of relativistic fluid is one of the main
postulates of Landau hydrodynamic theory \cite{La53} of multiparticle
production. We cannot prove or disprove this assumption in microscopic
simulations for the whole system, simply because a global
equilibrium is not attained. Though conditions in the cell are 
instantly changing, it is possible to check the behavior of the 
entropy per baryon. Within the 5\% accuracy limit, this ratio is 
nearly conserved in the equilibrium phase of the expansion, see
Fig.~\ref{s_vs_rhob}. The entropy densities obtained for the cell
in both models are very close to each other, but, because of the
difference in net-baryon sector, the ratio $s/\rho_{\rm B}$ in
UrQMD is about 15-20\% larger than that in QGSM. Together with the
pressure isotropy, the conservation of entropy per baryon supports 
the application of hydrodynamics. 

\begin{figure}[htb]
\includegraphics[scale=0.525]{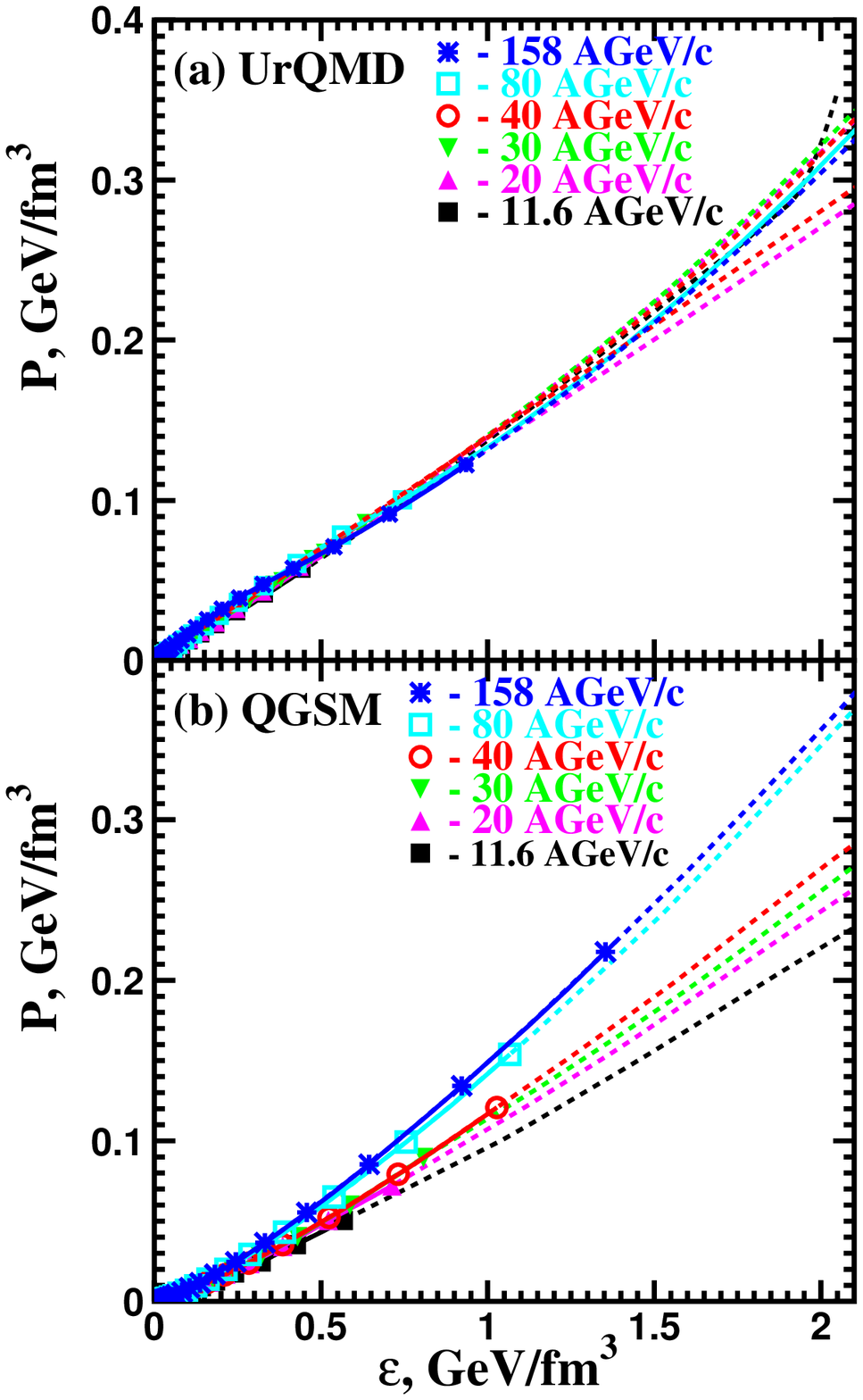}
\caption{
(Color online)
Time evolution of the microscopic pressure $P$ and the energy
density $\varepsilon$ in the central 125 fm$^3$ cell in (a) UrQMD and
(b) QGSM calculations of central Au+Au collisions at energies from
11.6{\it A} GeV to 158{\it A} GeV. Dashed lines correspond to the
non-equilibrium stage of the reaction, solid lines represent the
equilibrium phase.
\label{p_vs_e_mic}}
\end{figure}

\begin{figure}[htb]
\includegraphics[scale=0.525]{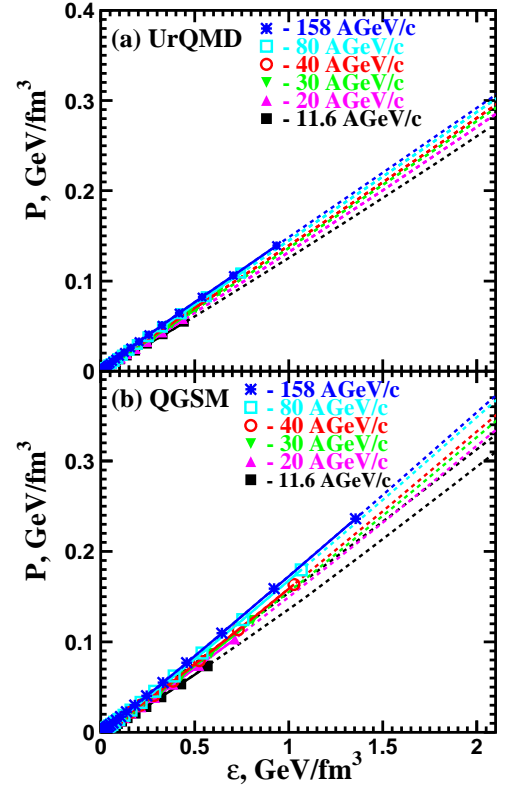}
\caption{
(Color online)
The same as Fig.~\protect\ref{p_vs_e_mic} but for
macroscopic pressure $P$ extracted from the SM fit to microscopic data.
\label{p_vs_e_sm}}
\end{figure}

Any hydrodynamic model relies on the equation of state, which links
the pressure of the system to its energy density. Otherwise, the 
system of hydrodynamic equations is incomplete. The corresponding 
plot with microscopic pressures $P_{mic}(\varepsilon)$ is presented in
Fig.~\ref{p_vs_e_mic}, whereas the macroscopic pressures obtained from
the SM fit are shown in Fig.~\ref{p_vs_e_sm}. In the last plot the
dependence of pressure on energy density is remarkably linear for both
models for all energies in question. Thus the EOS has a rather simple 
form
\beq
\ds
P(\varepsilon) = c_s^2 \varepsilon\ ,
\label{eq16}
\eeq
where the sonic velocity in the medium $c_s = (dP/d\varepsilon)^{1/2}$ 
is fully determined by the slopes of the distributions 
$P(\varepsilon)$. However, if the pressure is determined 
microscopically and not via the distribution function, the falloff of
pressure with decreasing energy density proceeds slightly nonlinearly.
This feature can be seen distinctly for top SPS energy in the QGSM
calculations. Therefore, for both models we averaged the slopes of 
the $P$ vs. $\varepsilon$ distributions over the whole period
of the equilibrated phase (see Fig.~\ref{p_vs_e_mic}). 
It should be noted that due to the averaging over time, respectively
energy density, the values do not represent the maximal values for 
$c_s^2$ which are reached in the corresponding reactions. They are 
actually lower, since also energy densities below the critical energy 
density of about $0.8\, {\rm GeV/fm^3}$ contribute to the average.

\begin{figure}[htb]
\includegraphics[scale=0.425]{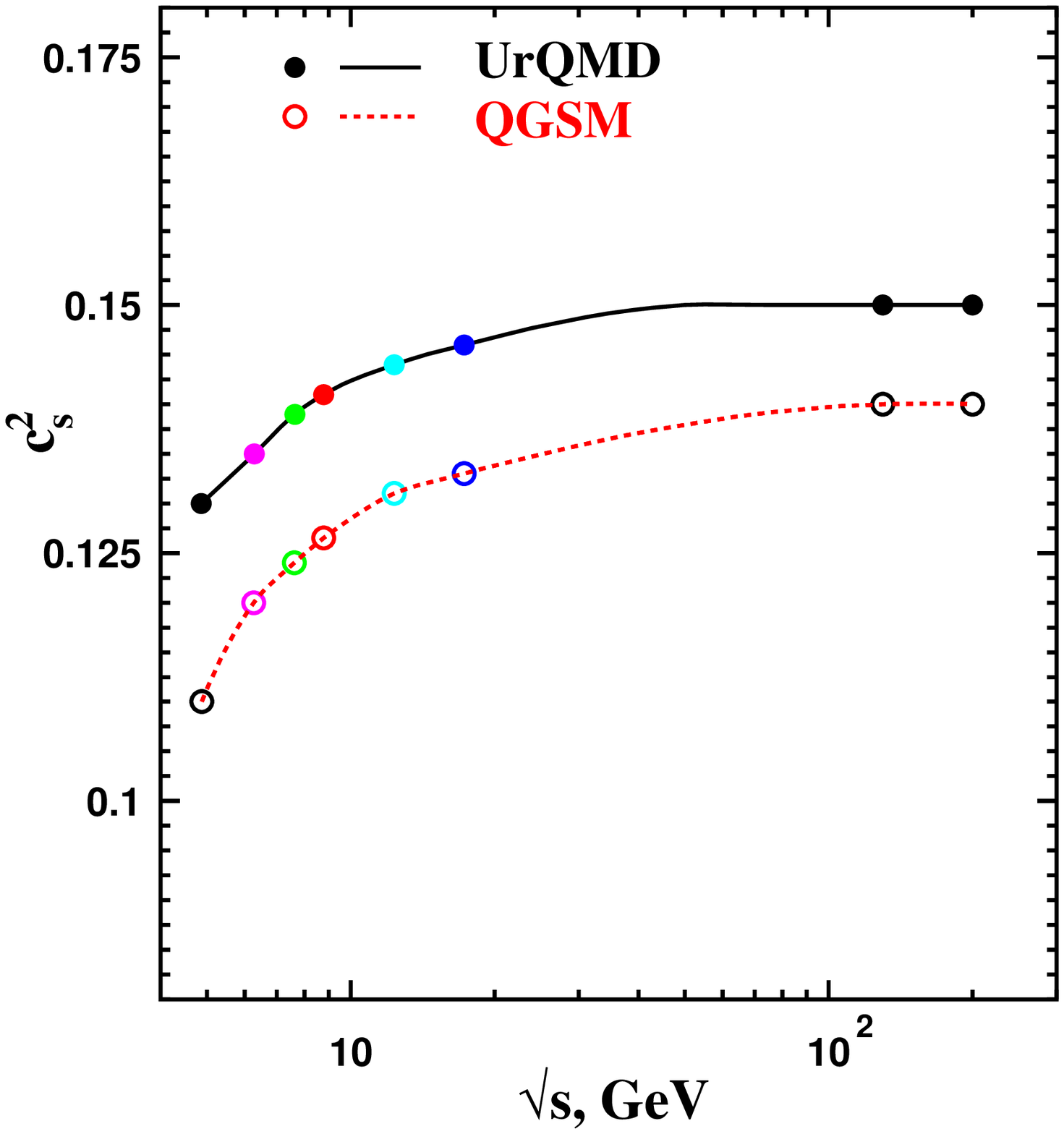}
\caption{
(Color online)
The ratio $P/\varepsilon = c_s^2$, where $P$ is defined
microscopically, in the central cell of volume
$V=125$\,fm$^3$ as a function of center of mass energy $\sqrt{s}$
in UrQMD (solid symbols) and in QGSM (open symbols) calculations.
Lines are drawn to guide the eye.
\label{cs2}}
\end{figure}

The extracted values of the $c_s^2$ are presented in Fig.~\ref{cs2}. 
For the UrQMD calculations the 
velocity of sound increases from 0.13 at $E_{lab} = 11.6${\it A} GeV 
to 0.146 at $E_{lab} = 158${\it A} GeV, and saturates at $c_s^2 = 
0.15$ for RHIC energies, $\sqrt{s} = 130${\it A} GeV and $\sqrt{s} = 
200${\it A} GeV \cite{prc01}. In QGSM calculations the averaged
sound velocity is about 0.015 units smaller due to the pion excess. 
For instance, it reaches 
$c_s^2 = 0.127$ at $E_{lab} = 40${\it A} GeV. Both models indicate
that at the energy around $E_{lab} = 40${\it A} GeV the slope of the
$c_s^2 (\sqrt{s})$ distribution is changing, and the velocity of
sound becomes less sensitive to rising bombarding energy.

Let us discuss the obtained values of the $c_s^2$. For the 
ultrarelativistic gas of light particles the well-known theoretical 
result is $c_s =1/\sqrt{3}$ of the speed of light \cite{LaLi80}. As 
shown in \cite{Shur72}, the presence of resonances in particle 
spectrum generates the decrease of the sonic speed. Employing the 
empirical dependence \cite{Hag65}
\beq
\ds
\rho (m) \propto m^{\alpha^\prime} \quad 
(2 \leq \alpha^\prime \leq 3)\ ,
\eeq
where  $\rho(m)\, dm$ denotes the number of resonances with masses from
$m$ to $m + dm$, one arrives to the equation of state in the form
\cite{Shur72}
\beq
\ds
\varepsilon = (\alpha^\prime + 4)\, P \ ,
\eeq
i.e., $\frac{1}{7} \leq c_s^2 \leq \frac{1}{6}$. This trend is 
reproduced in microscopic models. 

Since neither energy density nor pressure can be directly measured in
the central area of heavy-ion collisions, the experimental evaluation
of the $c_s$ is more difficult. One may rely on the hydrodynamic
calculations, which claim that the magnitude of the so-called elliptic
flow $v_2$ depends on the speed of sound $c_s$ \cite{BBBO05}. Using
the estimates, obtained for fixed impact parameter $b = 8$ fm under
assumption of constant $c_s$ throughout the system expansion, PHENIX
collaboration reported the value $c_s \approx 0.35 \pm 0.05$
\cite{PHENIX_Cs}, i.e. $c_s^2 \approx 0.12 \pm 0.3$, for gold-gold
collisions at top RHIC energy $\sqrt{s} = 200${\it A} GeV. This value
is close to our results and also implies rather soft effective EOS.  

Lattice calculations \cite{Aoki:2005vt}
predict an asymptotic value of $c_s^2 \sim 0.3$ slightly below the
Stefan-Boltzmann limit which indicates the appearance of a strongly
coupled partonic medium. Recombination processes decrease the mean free
path of the particles and lower thus the viscosity of the medium. By 
including such processes the sonic speed can be increased above the
critical energy density thus coming closer to the lattice predictions 
\cite{bleib_rec}.

\begin{figure}[htb]
\includegraphics[scale=0.525]{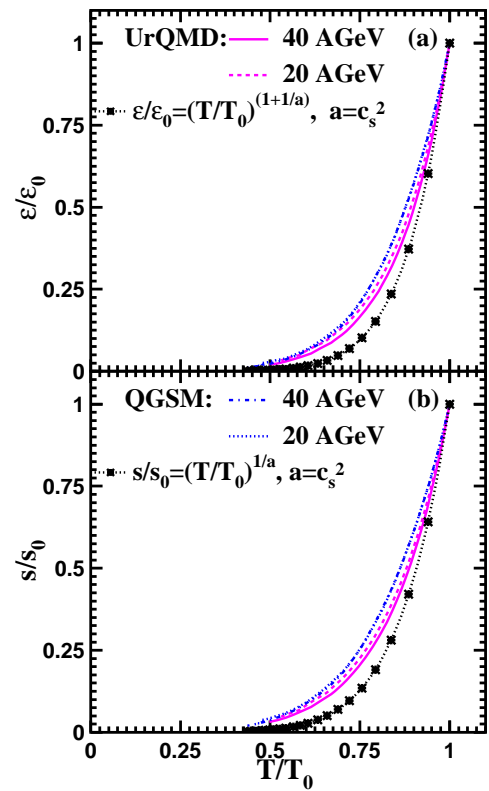}
\caption{
(Color online)
(a) Ratio of energy densities $\varepsilon/\varepsilon_0$ vs. $T/T_0$
in the central $V = 125$ fm$^3$ cell. Dashed line and solid line
represent UrQMD calculations of central Au+Au collisions at
20{\it A} GeV and 40{\it A} GeV, respectively, whereas dotted line
and dash-dotted line show the QGSM results for these reactions.
Asterisks depict the analytic calculations given by
Eqs.~(\protect\ref{eq17})-(\protect\ref{eq18}) with $\mu = 0$ and
$a = c_s^2 = 0.14$.
(b) The same as (a) but for the ratio of entropy densities $s/s_0$.
\label{en_entr_vs_t}}
\end{figure}

The velocity of sound defines the change of entropy and energy 
densities with decreasing temperature, provided the local 
equilibrium is maintained during the expansion. The analytic 
expressions, which can be derived, e.g., for gas of non-strange
mesons with zero chemical potential, read (see {\it Appendix\/}
\ref{app_B})
\beqar
\ds
\label{eq17}
\varepsilon &=& \varepsilon_0 \left( \frac{T}{T_0} \right)^\frac{
1+a}{a}\ ,\\
\label{eq18}
s &=& s_0 \left( \frac{T}{T_0} \right)^\frac{1}{a}\ .
\eeqar
The ratios $\varepsilon / \varepsilon_0$ and $s / s_0$ as functions
of $T / T_0$ obtained from model calculations at $E_{lab}=20${\it A} 
GeV and 40{\it A} GeV are plotted in Fig.~\ref{en_entr_vs_t}
together with results for $\mu = 0$ given by 
Eqs.~(\ref{eq17})-(\ref{eq18}). Although the hadron gas in the cell 
represents a cocktail of species with different chemical potentials, 
that can be either zero, positive or negative in case of 
antiparticles, the curves calculated by the UrQMD and QGSM are not far 
from the ideal ones. Moreover, there is just a very weak difference 
between the UrQMD and QGSM curves for both energies. If one formally 
fits these distributions to Eqs.~(\ref{eq17})-(\ref{eq18}) using the 
velocity of sound as fitting parameter, one gets $a = 0.2$ exactly. 
It would be nice to check whether the deceleration of energy(entropy) 
density falloff with dropping temperature could be charged solely to 
the presence of hadrons with non-zero chemical potential. One way to 
do this is to perform a similar analysis of the cell conditions at 
RHIC (or higher) energies. - Here strange hadrons, baryons and their 
resonances are still present \cite{prc01}, but both chemical 
potentials, $\mu_{\rm B}$ and $\mu_{\rm S}$, are quite small. 
Therefore, one may expect that the microscopic results would be 
closer to those presented by Eqs.~(\ref{eq17})-(\ref{eq18}).  

Note also, that pressure in the cell changes with
energy density quite smoothly, and no peculiarities which can be
attributed to first-order phase transition are seen in 
the early stage of the reaction. - Here we simply extend the formalism
of extraction of the thermodynamic parameters to the non-equilibrium
phase, where one cannot trust the obtained values anymore. This
was done merely in order to find any traces of the transition related 
to the onset of equilibrium and to changes of the effective EOS in
the models. However, the analysis is performed for the fixed cubic 
cell of relatively large volume $V = 125$\, fm$^3$, where the matter
is distributed non-homogeneously at early times. To get rid of the
evident ambiguities, the scheme is properly modified.

\subsection{Early stage of the evolution}
\label{subsec5d}

The central cell is further subdivided into the smaller ones,
embedded one into another. The size of the initial test-volume is 
just $V_{init} = 0.125$\, fm$^3$, and the energy density $\varepsilon$
of the cells becomes the main parameter now. If the $\varepsilon$ of
the inner cell is not the same (within the 5\% limit of accuracy) as 
the energy density of the outer one, the SM analysis of the 
thermodynamic conditions is performed for the inner cell. If the energy
density is uniformly distributed within the outer cell, the latter 
becomes a new test-volume, and so on.  
In the latter case it appears (see Fig.~\ref{s_vs_rhob}) that the
onset of the isentropic expansion regime in the central area occurs
significantly earlier than the formation of equilibrated matter.
Moreover, at the collision energies below 80{\it A} GeV entropy per 
baryon ratio seems to be quite stable almost from the beginning of
the reaction. 

\begin{figure}[htb]
\includegraphics[scale=0.525]{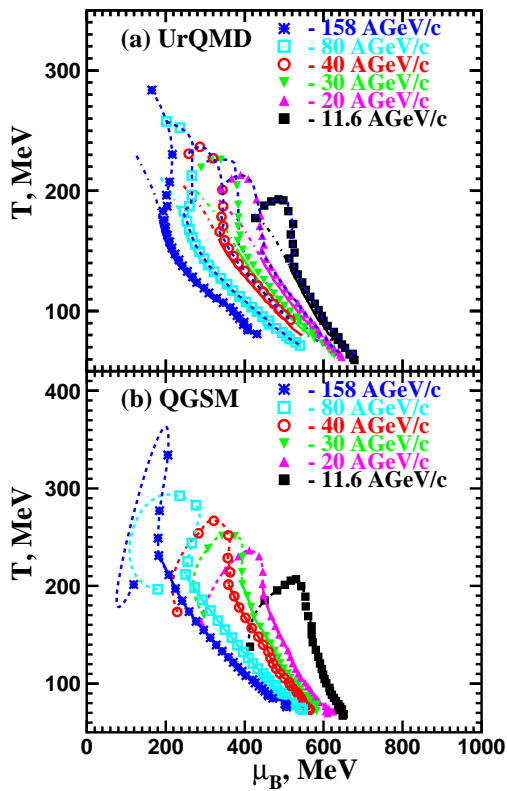}
\caption{
(Color online)
The evolution of the temperature $T$ and baryon chemical
potential $\mu_{\rm B}$ in the central cell of central Au+Au
collisions at energies from 11.6{\it A} GeV to 158{\it A} GeV. Both
parameters are extracted from the fit to the SM. Symbols and dashed
lines show the evolution of these quantities in a cell of instantly
increasing volume ($V_{init} = 0.125$\,fm$^3$), while dash-dotted
(upper plot) and full (both plots) lines are related to calculations
with the fixed volume $V = 125$\,fm$^3$.
\label{t_vs_mub}}
\end{figure}

Evolutions of the temperature and baryon chemical potential both in 
the central cell of the fixed volume $V = 125$\,fm$^3$ and in the 
expanding energy area are depicted in Fig.~\ref{t_vs_mub}. One sees

that the transition to equilibrium proceeds quite smoothly if the
analysis is performed for the fixed cell (Fig.~\ref{t_vs_mub}, upper 
plot). In contrast, in the area with uniformly distributed energy the
transition to the equilibrated phase is characterized by a kink 
distinctly seen in each of the phase diagrams in both microscopic 
models. Although this effect takes place along the lines of the
constant entropy per baryon, it should not be automatically linked to
the highly anticipated quark-hadron phase transition. The reason is
simple, - extraction of the thermodynamic parameters, such as $T, P,
\mu_{\rm B}$ and $\mu_{\rm S}$ (but not the entropy density, which is
determined microscopically), by means of the equilibrium statistical
model is doubtful for the non-equilibrium phase. On the other hand, 
the formation of the kink may not be accidental. It is correlated
with the significant reduction of the number of processes going via
the formation and fragmentation of strings, and, therefore, with
the inelastic (chemical) freeze-out of particles. - In both models 
the matter, produced in a central area in central heavy-ion 
collisions at energies between AGS and SPS, is dominated by 
(pseudo)elastic collisions after $t \approx 6 \div 8$ fm/$c$ 
\cite{BlAi02,fr_ags,fr_sps}. - In the fixed-cell analysis all 
parameters within the cell are averaged and the transition is smeared 
out. The observed phenomenon can easily mimic the signature of the 
QCD phase transition in the $T$-$\mu_{\rm B}$ plane, found in lattice 
QCD calculations \cite{eji06} also along the lines of the constant 
entropy per baryon.

Figure~\ref{t_vs_mub} demonstrates also that thermodynamic 
characteristics of the fixed-size cell and the instantly growing
energy-homogeneous area coincide completely during the equilibrium 
stage. In accord with earlier observation \cite{jpg99}, neither the 
mechanical reduction of the test-volume in longitudinal direction nor 
the criterion of uniformly distributed energy density alone can help 
us in searching for quick equilibration in the central zone of 
relativistic heavy-ion collisions. Criteria of local thermal and 
chemical equilibrium described in Sect.~\ref{sec4} are fulfilled after 
the chemical freeze-out in the test-volume, when the production of new 
particles in the system is ceased.

\section{Conclusions}
\label{sec6}

In summary, two different microscopic string models were used to
study the formation and evolution of the locally equilibrated
matter in the central zone of heavy-ion collisions at energies from
11.6{\it A} GeV to 160{\it A} GeV. Calculations were performed both
for the cubic central cell of fixed volume $V = 125$\, fm$^3$ and
for the instantly expanding area of homogeneous energy density.
Traditional approach based on the fulfillment of the conditions of
kinetic, thermal and chemical equilibrium has been applied to
decide whether or not the equilibrium is reached. Both models favor
the formation of the equilibrated matter for a period of about
10\, fm/$c$. During this period the expansion of matter in the 
central cell proceeds isentropically with constant entropy per
baryon.  The equation of state can be approximated by a simple 
linear dependence $P = a(\sqrt{s}) \varepsilon$, where the square of
the speed of sound $c_s^2 = a(\sqrt{s})$ varies from 0.13 (AGS) to 
0.15 (SPS) in the UrQMD calculations, and from 0.11 (AGS) to 0.14
(SPS) in the QGSM ones. 
In both models the rise of $a(\sqrt{s})$ with energy is slowed down 
after $E_{lab} = 40${\it A} GeV and saturates at RHIC energies. This 
change is assigned to the transition from baryon-dominated to 
meson-dominated matter.

Study of the expanding area of the isotropically distributed energy
reveals that the relaxation to equilibrium in this dynamic region
proceeds at the same rate as in the case of the fixed-size cell.
However, the entropy per baryon ratio becomes constant before the
state of equilibrium is attained. Here both microscopic models 
unambiguously show the presence of a kink in the $T$-$\mu_{\rm B}$
phase diagrams. The higher the collision energy, the earlier the
kink formation. Its origin is linked to the freeze-out of inelastic 
reactions in the considered area.

\begin{acknowledgments}
Fruitful discussions with N. Amelin, K. Boreskov, L. Csernai, 
A. Kaidalov, J. Randrup and L. Sarycheva are gratefully acknowledged.
We are especially indebted to late Nikolai Amelin, whom we dedicate
this paper.
L.B. and E.Z. are grateful to the Institute of Theoretical Physics,
University of T\"ubingen for the warm and kind hospitality.
This work was supported by the Norwegian Research Council (NFR)
under contract No. 166727/V30, the Deutsche Forschungsgemeinschaft
(DFG), and the Bundesministerium f{\"u}r Bildung und Forschung (BMBF) 
under contract 06T\"U986. 
\end{acknowledgments}

\appendix
\section{Reduction of temperature}
\label{app_A}

Let us consider non-relativistic ideal hadron gas which contains 
non-equilibrium number of pions, while the other hadron species 
correspond to their equilibrium values. In thermal equilibrium
the total energy of the gas is a sum of the masses of all particles
(potential term) and the energies of their thermal motion (kinetic
term). Compared to the case of fully equilibrized hadron gas, the
temperature of the system with overpopulated amount of pions should
reduce so that the total energies of both systems remain the same.

One can write
\beqar
\ds
\label{a1}
E^{(1)} &=& E^{(2)}\ ,\\
\label{a2}
E^{(1)} &=& \sum_{i} m_i^{(1)} N_i^{(1)} + \frac{3}{2} T^{(1)}
\sum_{i} N_i^{(1)}\ ,\\
\label{a3}
E^{(2)} &=& \sum_{i} m_i^{(2)} N_i^{(2)} + \frac{3}{2} T^{(2)}
\sum_{i} N_i^{(2)}\ ,\\
\label{a4}
N_{\pi}^{(1)} &=& \alpha N_{\pi}^{(2)}\ ,\\
\label{a5}
N_{i \neq \pi}^{(1)} &=& N_{i \neq \pi}^{(2)}\ ,
\eeqar
where the superscripts (1) and (2) are related to partially non-balanced
(w.r.t. pions) and fully equilibrized system, respectively. Parameter
$\alpha > 1$ measures the excess of pions in system (1). From 
Eqs.~(\ref{a1})-(\ref{a5}) we have
\beq
\ds
m_{\pi} N_{\pi}^{(2)} = \frac{3}{2} \left[ (T^{(2)} - T^{(1)} \sum_{i}
N_i^{(2)} - (\alpha - 1) T^{(1)} N_{\pi}^{(2)} \right]
\label{a6}
\eeq

Introducing the reduced variables
\beqar
\ds
\label{a7}
\beta &=& \sum_{i} N_i^{(2)} / N_{\pi}^{(2)}\ ,\quad (\beta > 1)\\   
\label{a8}
\gamma &=& T^{(1)} / T^{(2)}\ ,\quad (0 < \gamma < 1)\\
\label{a9}
\delta &=& \frac{2}{3} \frac{m_{\pi}}{T^{(2)}}\ ,
\eeqar

we get finally 

\beq
\beta (1 - \gamma) = (\alpha -1) (\gamma + \delta)
\label{a10}
\eeq

Now knowing the pion abundance in particle spectrum $\beta^{-1}$ at
chemical equilibrium and pion excess $\alpha$ one can estimate the
drop of temperature $\gamma$ in the system due to re-distribution of
kinetic energy among the extra degrees of freedom.

\section{Evolution of $\varepsilon$ and $s$ with $T$ at $\mu = 0$}
\label{app_B}

Gibbs free energy $G$ is linked to energy $E$ and entropy $S$ of 
the system with pressure $P$, volume $V$ and temperature $T$ via
the equality
\beq
\ds
G = E + PV -TS\ .
\label{b1}
\eeq
On the other hand, $G = \mu N$, where $\mu$ is the chemical
potential and $N$ is the number of particles. If the chemical
potential is absent, Eq.~(\ref{b1}) is reduced to the following
expression for the energy and entropy densities,
$\varepsilon = E/V$ and $s = S/V$, respectively:
\beq
\ds
\varepsilon +P = Ts
\label{b2}
\eeq
Utilizing the condition $\mu = 0$, one can derive from basic
thermodynamic equalities \cite{LaLi80}
\beqar
\ds
\label{b3}
d\varepsilon &=& T\, ds\ , \\
\label{b4}
dP &=& s\, dT\ .
\eeqar

Inserting the equation of state $dP = a\,d\varepsilon$ into these
equations, we get after straightforward calculations
\beqar
\ds
\label{b5}
\frac{a}{1+a} \frac{d\varepsilon}{\varepsilon} &=& \frac{dT}{T}\ ,\\
\label{b6}
a \frac{d s}{s} &=& \frac{dT}{T}\ ,
\eeqar
and, finally,
\beqar
\ds
\label{b7}
\frac{\varepsilon}{\varepsilon_0} &=&
\left( \frac{T}{T_0} \right)^\frac{1+a}{a}\ ,\\
\frac{s}{s_0} &=& \left( \frac{T}{T_0} \right)^\frac{1}{a}\ .
\eeqar

The obtained results are general for particles with $\mu = 0$ and
do not depend on the expansion, e.g. longitudinal or spherical,
scenario.

%%%%%%%%%%%%%%%%%%%%%%%%%%%%%%%%%%%%%%%%%%%%%%%%%%%%%%%%%%%%%%%%%%%
%%%%%%%%%%%%%%%%   REFERENCES  %%%%%%%%%%%%%%%%%%%%%%%%%%%%%%%%%%%%
%%%%%%%%%%%%%%%%%%%%%%%%%%%%%%%%%%%%%%%%%%%%%%%%%%%%%%%%%%%%%%%%%%%

\end{document}